\documentclass[11pt]{article}
\usepackage[dvipsnames]{xcolor}
\usepackage{tikz}
 \usetikzlibrary{positioning}
 \usetikzlibrary[patterns, patterns.meta]
 \usetikzlibrary[fadings]
 \usetikzlibrary{decorations.pathreplacing}
\usepackage{bm, bbm}
\usepackage[letterpaper, portrait, margin=1in]{geometry}
\usepackage[colorlinks=true,linkcolor=blue,citecolor=ForestGreen]{hyperref}
\usepackage{algorithm}
\usepackage[noend]{algpseudocode}
\usepackage{url}
\usepackage{amsmath,amssymb,amsthm}
\usepackage{thmtools,thm-restate}
\usepackage[noabbrev,capitalise,nameinlink]{cleveref}
\usepackage{mathtools}
\usepackage{xspace}
\usepackage{verbatim}
\usepackage{mathrsfs}
\usepackage{tabularx}
\usepackage{enumitem}
\usepackage{derivative}
\usepackage{multirow}
\usepackage{diagbox}
\usepackage{nicematrix}
\usepackage[most]{tcolorbox}
\usepackage{caption}

\usepackage{dsfont}

\newcommand*\ie{i.\kern.1em e.\ }
\newcommand*\eg{e.\kern.1em g.\ }
\newcommand*\cf{c.\kern.1em f.\ }
\newcommand*\almev{a.\kern.1em e.\ }

\definecolor{ama-iro}{RGB}{0, 158, 243.0}
\definecolor{fuyu-gaki}{RGB}{251, 74, 52}
\definecolor{momiji}{RGB}{245, 70, 111}
\definecolor{hotaru-bi}{RGB}{229,221,58} 
\definecolor{kon-peki}{RGB}{1,120,217}
\definecolor{shin-kai}{RGB}{77,98,152}
\definecolor{shin-ryoku}{RGB}{1,145,97}
\definecolor{yama-budo}{RGB}{171,14,122}

\definecolor{citecolor}{HTML}{5E9100}

\definecolor{theoremcolor}{HTML}{FFE1D9}
\definecolor{resultcolor}{HTML}{FFE1D9}

\definecolor{constraintcolor}{HTML}{BBD49B}
\definecolor{goalcolor}{HTML}{CAE4A7}

\definecolor{remarkcolor}{HTML}{E3EEC7}

\definecolor{definitioncolor}{HTML}{FCDFBE}

\definecolor{examplecolor}{HTML}{F9E9D9}
\definecolor{questioncolor}{HTML}{DDE8EB}

\hypersetup{
  colorlinks=true,
  linkcolor=momiji,
  filecolor=ama-iro,
  urlcolor=momiji,
  citecolor=citecolor
}

\definecolor{captioncolor}{RGB}{128,128,128}

\theoremstyle{plain}
\newtheorem{theorem}{Theorem}
\newtheorem{lemma}[theorem]{Lemma}

\newtheorem{proposition}[theorem]{Proposition}
\newtheorem{claim}[theorem]{Claim}
\newtheorem{corollary}[theorem]{Corollary}
\newtheorem{question}[theorem]{Question}
\newtheorem{question*}{Question}

\newtheorem{observation}[theorem]{Observation}
\newtheorem{assumption}[theorem]{Assumption}

\crefname{claim}{Claim}{Claims}
\crefname{fact}{Fact}{Facts}

\theoremstyle{definition}

\newtheorem{definition}[theorem]{Definition}
\newtheorem{remark}[theorem]{Remark}
\newtheorem{example}{Example}
\newtheorem{goal}[theorem]{Goal}
\newtheorem{goal*}{Goal}

\theoremstyle{plain}

\newenvironment{boxtheorem}{\begin{theorem}}{\end{theorem}}
\tcolorboxenvironment{boxtheorem}{colback=theoremcolor, colframe=white,
    colbacktitle=theoremcolor, coltitle=theoremcolor}

\newenvironment{boxlemma}{\begin{lemma}}{\end{lemma}}
\tcolorboxenvironment{boxlemma}{colback=resultcolor, colframe=white,
    colbacktitle=resultcolor, coltitle=resultcolor}

\tcolorboxenvironment{boxproposition}{colback=resultcolor, colframe=white,
    colbacktitle=resultcolor, coltitle=resultcolor}

\newenvironment{boxcorollary}{\begin{corollary}}{\end{corollary}}
\tcolorboxenvironment{boxcorollary}{colback=resultcolor, colframe=white,
    colbacktitle=resultcolor, coltitle=resultcolor}

\tcolorboxenvironment{boxobservation}{colback=resultcolor, colframe=white,
    colbacktitle=resultcolor, coltitle=resultcolor}

\tcolorboxenvironment{boxquestion}{colback=questioncolor, colframe=white,
    colbacktitle=questioncolor, coltitle=shin-kai}
\newenvironment{boxquestion*}{\begin{question*}}{\end{question*}}
\tcolorboxenvironment{boxquestion*}{colback=questioncolor, colframe=white,
    colbacktitle=questioncolor, coltitle=questioncolor}

\tcolorboxenvironment{boxdefinition}{colback=definitioncolor, colframe=white,
    colbacktitle=definitioncolor, coltitle=definitioncolor}

\tcolorboxenvironment{boxassumption}{colback=definitioncolor, colframe=white,
    colbacktitle=definitioncolor, coltitle=definitioncolor}

\tcolorboxenvironment{boxexample}{colback=examplecolor, colframe=white,
    colbacktitle=examplecolor, coltitle=examplecolor}

\newtheorem{exercise}[theorem]{Exercise}

\tcolorboxenvironment{boxexercise}{colback=exercisecolor, colframe=white,
    colbacktitle=exercisecolor, coltitle=exercisecolor}

\tcolorboxenvironment{boxgoal}{colback=goalcolor, colframe=white,
    colbacktitle=goalcolor, coltitle=goalcolor}

\newenvironment{boxgoal*}{\begin{goal*}}{\end{goal*}}
\tcolorboxenvironment{boxgoal*}{colback=goalcolor, colframe=white,
    colbacktitle=goalcolor, coltitle=goalcolor}

\tcolorboxenvironment{boxremark}{colback=remarkcolor, colframe=white,
    colbacktitle=remarkcolor, coltitle=remarkcolor}

\theoremstyle{plain}

\newcounter{constraintcounter}
\renewcommand{\theconstraintcounter}{%
  \ifnum\value{constraintcounter}=5 IVb\else \Roman{constraintcounter}\fi
}

\newtheorem{constraint}[constraintcounter]{Constraint}
\newtheorem{constraint*}{Constraint}
\crefname{constraint}{Constraint}{Constraints}

\tcolorboxenvironment{boxconstraint}{colback=constraintcolor, colframe=white,
    colbacktitle=constraintcolor, coltitle=constraintcolor}

\newenvironment{boxconstraint*}{\begin{constraint*}}{\end{constraint*}}
\tcolorboxenvironment{boxconstraint*}{colback=constraintcolor, colframe=white,
    colbacktitle=constraintcolor, coltitle=constraintcolor}

\newcommand{\ignore}[1]{}

\DeclareMathOperator{\supp}{supp}   
\DeclareMathOperator{\poly}{poly}

\newcommand*{\define}{\mathrel{\vcenter{\baselineskip0.5ex \lineskiplimit0pt
                      \hbox{\scriptsize.}\hbox{\scriptsize.}}}%
                      =}

\newcommand{\zo}{\{0,1\}}

\newcommand{\cP}{\ensuremath{\mathcal{P}}}
\newcommand{\cQ}{\ensuremath{\mathcal{Q}}}

\newcommand{\cZ}{\ensuremath{\mathcal{Z}}}

\newcommand{\bF}{\ensuremath{\mathbb{F}}}
\newcommand{\bN}{\ensuremath{\mathbb{N}}}

\newcommand{\bR}{\ensuremath{\mathbb{R}}}

\newcommand{\tOmega}{\tilde{\Omega}} 

\newcommand{\BPP}{\mathsf{BPP}}

\newcommand{\GT}{\textsc{GT}}
\renewcommand{\setminus}{\smallsetminus}

\algnewcommand\algorithmicinput{\textbf{Input: }}
\algnewcommand\Input{\item[\algorithmicinput]}
\algnewcommand\algorithmicoutput{\textbf{Output: }}
\algnewcommand\Output{\item[\algorithmicoutput]}
\algnewcommand{\OneLineIf}[2]{
  \State \algorithmicif\ #1\ \algorithmicthen\ #2}

\definecolor{CommentColor}{RGB}{102,102,102}

\newtoggle{random}
\togglefalse{random}

\title{Equality is Far Weaker than Constant-Cost Communication}
\author{
Mika G\"o\"os \\[-1pt]
{\normalsize\slshape EPFL}\\[-1pt]
\iftoggle{random}{\texttt{mika.goos@epfl.ch}}
\and
Nathaniel Harms \\[-1pt]
{\normalsize\slshape EPFL \& UBC}\\[-1pt]
\iftoggle{random}{\texttt{nharms@cs.ubc.ca}}
\and
Artur Riazanov \\[-1pt]
{\normalsize\slshape EPFL}\\[-1pt]
\iftoggle{random}{\texttt{tunyash@gmail.com}}
}

\date{}

\begin{document}

\maketitle

\newcommand{\Eq}{\textsc{Eq}}
\newcommand{\R}{\mathsf{R}}
\newcommand{\RPDT}{\mathsf{RPDT}}
\newcommand{\D}{\mathsf{D}}
\newcommand{\N}{\mathsf{N}}
\newcommand{\bc}{\mathrm{bc}}

\newcommand{\Rpriv}{\mathsf{R}^{\mathrm{priv}}}
\newcommand{\DEQ}{\mathsf{D}^{\textsc{Eq}}}
\newcommand{\NEQ}{\mathsf{NP}^{\textsc{Eq}}}
\newcommand{\HD}[2]{\textsc{HD}_{#1}^{#2}}
\newcommand{\IIP}{\textsc{IIP}}
\newcommand{\EQ}{\textsc{Eq}}
\newcommand{\tr}{\mathrm{tr}}
\newcommand{\rkone}{\textsc{RankOne}}
\newcommand{\rk}{\mathrm{rk}}
\renewcommand{\H}{\mathrm{H}}
\renewcommand{\P}{\mathsf{P}}
\newcommand{\NP}{\mathsf{NP}}
\newcommand{\ND}{\mathsf{ND}}

\maketitle

\begin{abstract}\noindent
We exhibit an $n$-bit communication problem with a constant-cost randomized
protocol but which requires $n^{\Omega(1)}$ deterministic (or even
non-deterministic) queries to an \textsc{Equality} oracle. Therefore, even
\emph{constant-cost} randomized protocols cannot be efficiently ``derandomized''
using \textsc{Equality} oracles. This improves on several recent results and
answers a question from the survey of Hatami and Hatami~({\footnotesize SIGACT
News 2024}). It also gives a significantly simpler and quantitatively superior
proof of the main result of Fang, G\"o\"os, Harms, and Hatami~({\footnotesize
STOC~2025}), that constant-cost communication does not reduce to the
\textsc{$k$-Hamming Distance} hierarchy.
\end{abstract}

\iftoggle{random}{%
\thispagestyle{empty}
\newpage
\setcounter{page}{1}
}

\section{Introduction}

In this paper we prove strong limits on the power of ``simple
hashing'' for the purpose of communication. In communication complexity, simple
hashing, \ie the \textsc{Equality} communication problem, is the most dramatic
example of the power of randomness. Alice and Bob each have binary
strings~$x,y \in \zo^n$ and their goal is to decide whether $x=y$. With 
public randomness, they can generate a random 2-bit hash $\bm h(z)$ for every
$z$ and check if $\bm h(x) = \bm h(y)$. This succeeds with probability at least
$3/4$ and requires only 2 bits of communication regardless of the size of the
input $n$ \cite{KN96,RY20}.

Randomized communication can often be ``derandomized'' by allowing Alice and Bob
access to an oracle which (deterministically) computes \textsc{Equality}. Write
$\DEQ(F)$ for the minimum cost of a deterministic protocol computing $F$ with
access to this oracle, and $\R(F)$ for the public-coin randomized communication cost (with error probability $1/3$). Here are three examples covering a range of complexities:

\begin{example}[Greater Than]
Alice and Bob are
given $n$-bit integers $x, y$ and their goal is to decide whether $x > y$. They
can perform binary search to find the highest-order bit where $x$ and $y$
differ, by querying the \textsc{Equality} oracle (check if the first half of
their bits are equal, etc.), so $\DEQ(\GT_n) = O(\log n)$, where $\GT_n$ denotes
the \textsc{Greater-Than} problem on $n$ bits. Randomized communication 
satisfies $\R(F) = O(\DEQ(F))$ \cite{Nis93,HR24}, which in this case
gives the optimal randomized protocol: $\R(\GT_n) = \Theta(\log n)$
\cite{Nis93,BW16,Vio15,SA23}.
\end{example}

\begin{example}[Planar Adjacency \cite{Har20,HWZ22}]
Alice and Bob are given vertices $x,y$ in a (shared) planar graph $P$ and wish
to decide if they are adjacent. A planar graph is the edge union of 3 forests, so
they can check adjacency in $P$ by checking adjacency in each forest: $x$ is
equal to the parent of $y$ or vice versa. So $\DEQ \leq 6$ for this problem and
therefore its randomized cost is \emph{constant} (independent of the input
size), which is clearly optimal.
\end{example}

\begin{example}[1-Hamming Distance]
\label{ex:intro-1hd}
Alice and Bob are given $x,y \in \zo^n$ and their goal is to decide if $x,y$
differ on at most 1 bit. Using binary search with the
\textsc{Equality} oracle, we get $\DEQ(\HD{1}{n}) = O(\log n)$, where $\HD{1}{n}$
denotes the \textsc{1-Hamming Distance} problem on $n$ bits. This
protocol is \emph{not} optimal, since $\R(\HD{1}{n}) =
O(1)$, but it is still \emph{efficient}, \ie $\DEQ(\HD{1}{n}) = \poly\log n$.
\end{example}

Can \emph{every} efficient randomized protocol be replaced
with an \textsc{Equality} oracle protocol? This is especially interesting for
problems like \textsc{Planar Adjacency} and \textsc{1-Hamming Distance} that
have \emph{constant} randomized cost, because one may reasonably expect the
answer to be \emph{yes}, \ie any constant-cost randomized protocol can be replaced with
$\poly\log n$ \textsc{Equality} queries, as in \cref{ex:intro-1hd}. On the
contrary, we show:

\begin{boxtheorem}\label{thm:main}
There exists a communication problem $F\colon\{0,1\}^n\times\{0,1\}^n\to\{0,1\}$ with
\[
\R(F) = O(1)
\qquad\text{and}\qquad
\D^\Eq(F) =  \Theta(\sqrt{n}).
\]
\end{boxtheorem}
In terms of complexity classes, this implies $\BPP_0 \not\subseteq \P^\EQ$,
where $\BPP_0$ is the class of randomized constant-cost problems and $\P^\EQ$ is
the class of problems with \textsc{Equality} oracle cost $\poly\log n$.

Communication oracles including \textsc{Equality}, $\HD{1}{}$,
and others, are well studied
\cite{CLV19,HHH23,HWZ22,EHK22,PSS23,CHHS23,CHZZ24,HZ24,FHHH24,FGHH25,CHHNPS25}.
\cref{thm:main} improves on results from several of these works and our proof
answers a number of open questions.

\paragraph*{Separation of $\R(F)$ and $\DEQ(F)$.}
Whether \textsc{Equality} oracles can efficiently derandomize communication was
first asked in \cite{CLV19}, who   showed that there exists a problem $F$
with
\[
\R(F) = O(\log n)
\qquad\text{and}\qquad
\D^\Eq(F) =  \Theta(n) .
\]
This established that $\BPP
\not\subseteq \P^\EQ$, where $\BPP$ is the class of $n$-bit communication
problems with randomized communication cost $\poly\log n$. This separation was
improved in \cite{CHHNPS25} to hold for XOR functions (see also
\cite{CHHS23,PSS23,Tom25}). 

Recently, there has been significant effort to understand the most extreme
examples of the power of randomness in communication, \ie the problems with
constant cost. The question of how well \textsc{Equality} oracles can
derandomize constant cost communication was posed in \cite{HHH23,HWZ22}, who
showed that \textsc{Equality} is not ``complete'' for the class of constant-cost
problems, \ie $\BPP_0 \not\subseteq \P_0^\EQ$, where $\P_0^\EQ$ is the class of
problems $F$ with $\DEQ(F) = O(1)$. But the best known lower bound on $\DEQ$ for
a constant-cost problem is from \cite{HHH23} (see also \cite{HR24}), who proved
a $\DEQ$ lower bound matching \cref{ex:intro-1hd}:
\[
\R(\HD{1}{}) = O(1)
\qquad\text{and}\qquad
\D^\Eq(\HD{1}{}) =  \Theta(\log n) .
\]
This leads to the question, implicit in \cite{CHHS23} and explicit in the survey
\cite{HH24}, of whether constant-cost protocols can be replaced with $\poly\log
n$ \textsc{Equality} queries. We answer \emph{no}, improving the
separation $\BPP \not\subseteq \P^\EQ$ to $\BPP_0 \not\subseteq \P^\EQ$. Our
example $F$ is an XOR function, so we also improve on the separation of
\cite{CHHNPS25}.

\paragraph*{Separation of $\BPP_0$ and \textsc{$k$-Hamming Distance}.}

Following the result that \textsc{Equality} is not complete for $\BPP_0$, \cite{FGHH25} showed that the infinite hierarchy of \textsc{$k$-Hamming
Distance} problems $\HD{k}{}$ (decide if $x,y$ differ on at most $k$ bits) is
also not complete: there exists $F$ with $\R(F) = O(1)$ and yet
$\D^{\HD{k}{}}(F) = \omega(1)$ for every constant $k$, where $\D^{\HD{k}{}}(F)$
is the minimum number of queries to an $\HD{k}{}$ oracle required to compute
$F$. We give a significantly simpler and quantitatively superior proof (for a
different $F$), with a lower bound of $\tOmega(\sqrt n)$ instead of
$\omega(1)$:

\begin{boxcorollary}\label{cor:main}
There exists an $n$-bit function $F$ such that for any constant $k$,
\[
\R(F) = O(1)
\qquad\text{and}\qquad
\D^{\textsc{HD}_k}(F)\geq \tOmega(\sqrt{n}).
\]
\end{boxcorollary}

\subsection*{Two Proofs with Five Corollaries}

We give two incomparable proofs of \cref{thm:main}: one analytic, one
combinatorial, each with different consequences beyond the ones above, which we elaborate
in \cref{section:consequences}.

\paragraph*{Analytic proof: $\gamma_2$-norm and decision trees.}
We show in \cref{thm:gamma2} that the
\emph{$\gamma_2$-norm} of our problem~$F$ (equivalently, the spectral norm of
$f$, see \cref{section:pdt}) is $2^{\Omega(\sqrt n)}$. As a consequence:
\begin{itemize}
\item Our function $F$ is an XOR function, meaning $F(x,y) = f(x \oplus y)$ for some
function $f$. Since XOR functions $F(x,y) = f(x \oplus y)$ satisfy the property
that the $\gamma_2$-norm of $F$ is equal to the Fourier spectral norm of $f$
\cite{HHH23}, we exhibit a function $f$ with \emph{approximate} spectral norm
$O(1)$ yet \emph{exact} spectral norm $2^{\Omega(\sqrt n)}$
(\cref{cor:approx-spectral-norm}).
\item We answer \cite[Question 8]{CHHNPS25}, asking whether there exists a function $f \colon \zo^n \to \zo$
with randomized parity decision tree size $O(1)$ but deterministic parity
decision tree size $n^{\omega(1)}$. Our
$f$ has deterministic parity decision tree size $2^{\Omega(\sqrt n)}$ (\cref{cor:rpdt}).
\item We make progress on \cite[Question
5]{CHHS23}, asking whether any $n$-bit function with~$\R(F) = O(1)$ has
$\gamma_2$-norm $2^{\Omega(n)}$. Our function does not achieve this maximal
value but it improves exponentially on the best known value of $\Theta(\sqrt n)$,
for the $\HD{1}{}$ function \cite{HHH23}. 
\end{itemize}

\paragraph*{Combinatorial proof: non-deterministic \textsc{Equality} protocols.}
Using the \emph{blocky cover number} of \cite{PSS23},
we show that $F$ has \textsc{Equality} oracle cost $\Omega(\sqrt n)$ even for a
\emph{non-deterministic} protocol. As a consequence of our proof:
\begin{itemize}
\item We improve the complexity class separation in \cref{thm:main}, from
$\BPP_0 \not\subseteq \P^\EQ$ to $\BPP_0 \not\subseteq \NP^\EQ$. This improves 
on a result of \cite{PSS23} who showed $\BPP \not\subseteq \NP^\EQ$.
\item We tighten the theorem of \cite{PSS23} relating the \emph{blocky cover number} and
the non-deterministic \textsc{Equality} oracle cost $\ND^\EQ(\cdot )$. This
improves the lower bounds on $\ND^\EQ(\cdot)$ given in \cite{PSS23,CHHNPS25} and
answers the question posed after \cite[Theorem 5]{CHHNPS25}, asking
if $\ND^\EQ(\IIP_3^n) = \Omega(n)$, where
$\IIP_3^n$ is the \textsc{Integer Inner Product} function defined in
\cite{CLV19}.
\end{itemize}

\section{The Problem}

The problem is a special case of those studied recently by Sherstov \&
Storozhenko \cite{SS24}. Alice and Bob are given matrices $A, B \in \bF_2^{n
\times n}$ and their goal is to decide if the $\bF_2$-rank of $A + B$ is at most
1; \ie $\rkone_n \colon \zo^{n \times n} \times \zo^{n \times n} \to \zo$ is
defined by
\[
    \rkone_n(A,B) \define \begin{cases}
        1 &\text{ if } A \oplus B \text{ has rank $\leq 1$} \\
        0 &\text{ otherwise.}
    \end{cases}
\]
Note that a matrix $C \in \zo^{n \times n}$ has rank $\leq 1$ if and only if its
1-entries form a combinatorial rectangle, \ie there are sets $X, Y \subseteq
[n]$ such that $C(x,y) = 1$ iff $(x,y) \in X \times Y$.
This problem is an XOR problem, \ie $\rkone_n(A,B) = f(A \oplus B)$ for the
boolean function $f \colon \zo^{n \times n} \to \zo$ defined as $f(M) = 1$ if and only
if $M$ has rank at most $1$.

Sherstov \& Storozhenko \cite{SS24} prove tight bounds for the randomized
communication complexity of computing rank $r$ over any finite field. For the
sake of completeness, we include a simple proof that $\R(\rkone_n) = O(1)$. More
specifically, we prove that the randomized parity decision tree depth of $f$ is
$O(1)$ (which immediately implies the same for randomized
communication).

A function $g\colon \zo^n \to \zo$ is computed by a depth-$d$ parity
decision tree if, for any $x \in \zo^n$, $g(x)$ can be computed with at most $d$
adaptive queries of the form $\oplus_{i \in S} x_i$ for chosen $S \subseteq
[n]$. A function $g$ is computed by a \emph{randomized} parity decision tree if
there exists a distribution $\mu$ over parity decision trees such that for every
input $x \in \{0,1\}^n$, $\Pr_{\bm{T} \sim \mu}[\bm{T}(x) = g(x)] \ge 2/3$. 

\begin{theorem}
\label{thm:rpdt-upper-bound}
    Let $f\colon \{0,1\}^{n \times n} \to \{0,1\}$ be as above: $f(M) = 1$ iff $\rk_{\mathbb{F}_2}(M) \le 1$. Then it is computed by a randomized parity decision
    tree of constant depth (\ie independent of $n$).
\end{theorem}
\begin{proof}
    Given input $M$, the tree chooses sets $\bm{A}_1, \bm{A}_2, \bm{B}_1, \bm{B}_2 \subseteq [n]$
    uniformly at random.  For each  $\alpha, \beta \in [2]$, the tree queries
    $\bm{C}_{\alpha,\beta} \define \bigoplus_{(i,j) \in \bm{A}_\alpha \times
    \bm{B}_\beta} M_{ij}$, to form the $2 \times 2$ matrix $\bm C$. The tree
    outputs $1$ iff $\rk(\bm{C}) \le 1$. 

    The transformation from $M$ to $\bm{C}$ can be seen as a two-step process:
    first, we generate a $n \times 2$ matrix $\bm{C}'$ where column $i \in [2]$
    is the sum of columns $j \in \bm{B}_i$ of $M$. Then we generate the $2
    \times 2$ matrix $\bm{C}$ where row $j$ is the sum of rows $i \in \bm{A}_j$
    of $\bm{C}'$. Neither step may increase the rank of the matrix,  so if
    $\rk(M) \le 1$, then $\Pr[\rk(\bm{C}) \le 1] = 1$.
    
    Now suppose $\rk(M) \ge 2$. Each column of $\bm{C}'$ is chosen uniformly
    from the column space of $M$. The size of the column space is at least $4$,
    so the probability that both columns of $\bm{C}'$ are nonzero and distinct
    is at least $\tfrac{3}{4}\cdot\tfrac{1}{2} = \tfrac{3}{8}$. Conditional on
    this event, by the same argument, the probability that both rows of $\bm C$
    are nonzero and distinct is at least $\tfrac{3}{8}$. Therefore, the tree
    will output 0 with probability at least $9/64$. We can boost this to $2/3$
    by repetition.
\end{proof}

To clarify notation, observe that the inputs to this function are naturally expressed with $n^2$
variables, whereas \cref{thm:main} is stated for a function on $n$
variables. So to prove \cref{thm:main} we must show 
\begin{equation}
    \label{eq:relabel-variables}
    \DEQ(\rkone_n) = \Theta(n) .
\end{equation}
The upper bound is simple:
\begin{proposition} \label{prop:deq-ub}
     $\DEQ(\rkone_n) = O(n)$.
\end{proposition}
\begin{proof}
    Suppose Alice receives a matrix $A$ with rows $a_1, \dots, a_n \in
    \{0,1\}^n$, and Bob receives a matrix~$B$ with rows $b_1, \dots, b_n \in
    \{0,1\}^n$. Then $\rk(A \oplus B) \le 1$ if and only if each pair $i,j \in [n]$
    with $a_i \neq b_i$ and $a_j \neq b_j$ satisfy $a_i \oplus b_i = a_j \oplus
    b_j$. Using $n$ queries $\EQ(a_1, b_1), \EQ(a_2, b_2), \dots, \EQ(a_n,
    b_n)$, Alice and Bob can find $i$ such that $a_i \neq b_i$ if such an $i$
    exists (otherwise $\rk(A \oplus B) = 0$). Then for every $j \in [n]$ such
    that $a_j \neq b_j$ they check that $a_j \oplus b_j = a_i \oplus b_i$ by
    making the query $\EQ(a_j \oplus a_i, b_j \oplus b_i)$. If any
    \textsc{Equality}-query comes out negative, Alice and Bob output $0$,
    otherwise they output $1$.
\end{proof}

\section{Analytic Proof}

We analyze the $\gamma_2$-norm of $\rkone_n$. The $\gamma_2$-norm of a matrix $M \in \bR^{N \times N}$
is defined as
\[
    \gamma_2(M) \define \min_{M=UV} \|U\|_\text{row} \|V\|_\text{col} ,
\]
where the minimum is over matrices $U,V \in \bR^d$ (in any dimension $d$)
satisfying $M = UV$, $\|U\|_\text{row}$ is the maximum $\ell_2$-norm of any row
of $U$, and $\|V\|_\text{col}$ is the maximum $\ell_2$-norm of any column
of~$V$. The $\gamma_2$-norm is related to the \textsc{Equality} oracle cost by
\begin{equation}
    \label{eq:gamma2-deq}
    \forall M \in \zo^{N \times N} \;\colon\qquad \frac{1}{2} \log \gamma_2(M) \leq \DEQ(M) \qquad\text{(\cite{HHH23})}.
\end{equation}
To give a lower bound on $\gamma_2$, we use the H\"older's inequality framework
introduced in \cite{CHHNPS25}. Write $\|M\|_F \define (\sum_{i,j \in [N]} M_{i,j}^2)^{1/2}$
for the Frobenius norm of a matrix $M \in \bR^{N \times N}$. Then:
\begin{lemma}[Corollary of H\"older's inequality, \cite{CHHNPS25}]
\label{lem:holder}
    For any matrix $M \in \mathbb{R}^{N \times N}$,
    \[ \gamma_2(M) \ge \frac{1}{N} \cdot \frac{\|M\|^3_F}{\sqrt{\tr((M^T M)^2})}. \]
\end{lemma}

In the next theorem, note that the number of bits in the input of $\rkone_n$ is
$n^2$, not $n$; by renaming the number of bits and applying \cref{eq:gamma2-deq},
we get the bound in \cref{thm:main}.

\begin{boxtheorem}
\label{thm:gamma2}
    $\gamma_2(\rkone_n) = 2^{\Theta(n)}$. As a consequence, $\DEQ(\rkone_n) = \Omega(n)$.
\end{boxtheorem}
\begin{proof}
The upper bound follows from \cref{prop:deq-ub} and \cref{eq:gamma2-deq}. For the lower bound, we shall apply \cref{lem:holder} to the matrix $\rkone_n \in \{0,1\}^{N \times N}$ where $N = 2^{n^2}$.
    For convenience we write $M \define \rkone_n$.
    Since this matrix is boolean, the Frobenius norm is $\|M\|_F =
    (\sum_{i,j \in [N]} M^2_{i,j})^{1/2} = (\sum_{i,j \in [N]} M_{i,j})^{1/2}$, so
    it suffices to count the number of 1-valued entries. Rows of $M$ are
    identified with matrices $A \in \zo^{n \times n}$. There are $2^{2n}$
    rectangles $R \subseteq [n] \times [n]$ and they are in 1-to-1 correspondence with rank-1 matrices $Z \in \zo^{n \times n}$; here and in what follows, let us abuse language and consider the all-0 matrix (corresponding to the empty rectangle) to also be ``rank-1''. For fixed row $A$, each rank-1 matrix $Z$ has a unique $B \in \zo^{n \times n}$ such that $A \oplus B = Z$. Therefore the number of 1-valued entries in each row $A$ is $2^{2n}$. So
    \begin{equation}
    \label{eq:frobenius}
        \|M\|_F = \sqrt{N \cdot 2^{2n}} = 2^n \cdot \sqrt N .
    \end{equation}
    Now we bound the denominator in \cref{lem:holder},
    \[
        \tr((M^T M)^2) = \sum_{x,y,z,w \in [N]} M_{x,z} M_{x,w} M_{y,z} M_{y,w} .
    \]
    Rows and columns $x,y,z,w \in [N]$ are identified with matrices $X,Y,Z,W
    \in \{0,1\}^{n \times n}$. The product $M_{x,z} M_{x,t} M_{y,z} M_{y,w}$
    equals $1$ iff $R_1 \define X \oplus Z$, $R_2 \define X \oplus W $,
    $R_3 \define Y \oplus Z$, and $R_4 \define Y \oplus W$ are all rank-$1$ matrices in $\{0,1\}^{n
    \times n}$. Then the quadruple $(X,Y,Z,W)$ is uniquely determined by $(X,
    R_1, \dots, R_4)$ which in turn is determined by $(X, R_1, R_2, R_3)$ since
    $R_4 = R_1 \oplus R_2 \oplus R_3$. So:
    \begin{align*}
        \sum_{x,y,z,w} M_{x,z} M_{x,t} M_{y,z} M_{y,w}
            &= N \cdot \big|\{(R_1,R_2,R_3) \in (\{0,1\}^{n \times n})^3 \mid \rk(R_i) \leq 1;\; \rk(R_1 \oplus R_2 \oplus R_3) \leq 1\}\big|.
    \end{align*}
    Now it remains to bound the number of triples $(R_1, R_2, R_3)$ of rank-$1$ matrices that
    sum to a rank-$1$ matrix. Recall that rank-$1$ matrices over $\mathbb{F}_2$ are
    precisely the matrices whose $1$-entries form a combinatorial rectangle, let
    $A_i \times B_i \subseteq [n]^2$ be this rectangle for $R_i$ for $i \in
    [3]$. 

    We say that two sets $A,B$ are in \emph{general position} if they are not
    disjoint and neither set is a subset of another. In other words $A\setminus
    B$, $A \cap B$, and $B \setminus A$ are all non-empty. We first show that
    there are at most $6 \cdot 3^n \cdot 2^{4n}$ triples $(R_1, R_2, R_3)$
    where the sides of $R_1, R_2$ are \emph{not} in general position. Then we
    show that, if the sides of the first two rectangles $R_1, R_2$ \emph{are} in
    general position, there is only a constant number of choices for $R_3$, meaning that
    there are at most $9 \cdot 2^{4n}$ of these triples. Therefore
    \[
        \tr((M^T M)^2) = \sum_{x,y,z,w} M_{x,z} M_{x,t} M_{y,z} M_{y,w}
            \leq N \left( 6 \cdot 3^n \cdot 2^{4n} + 9 \cdot 2^{4n} \right) \leq 15 \cdot N \cdot 3^n \cdot 2^{4n} .
    \]
    Then \cref{lem:holder} implies
    \[
        \gamma_2(\rkone_n) \geq \frac{1}{N} \cdot \frac{N^{3/2} 2^{3n}}{\sqrt{15} \cdot \sqrt{N} \cdot (3 \cdot 2^4)^{n/2}}
        = \frac{1}{\sqrt{15} } \cdot \left(\frac{8}{\sqrt{3} \cdot 4}\right)^n = 2^{\Omega(n)} ,
    \]
    which concludes the proof, once we establish the claimed bounds on the number of triples.
    
    \paragraph{Structured pairs.} We count the number of triples $(R_1, R_2, R_3)$ where $R_1 = A_1 \times B_1$ and $R_2 = A_2 \times B_2$ have either $A_1, A_2$ or $B_1, B_2$ not in general position. First we count
    the number of choices for $A_1, A_2$ where $A_1 \subseteq A_2$. This is at most $3^n$ because each $i \in [n]$ can be included in $A_2\setminus A_1$, $A_1$, or neither. Therefore the number of triples $(R_1, R_2, R_3)$ where $A_1 \subseteq A_2$ is at most $3^n \cdot (2^n)^{4}$, accounting for the choices of $A_3, B_1, B_2, B_3$. Similar arguments hold for the other cases of non-general position of $A_1, A_2$ or $B_1, B_2$. There are 6 cases in total, so the number of triples $(R_1, R_2, R_3)$ is at most $6 \cdot 3^n \cdot 2^{4n}$.
    
    \paragraph{General position.} We count the number of triples $(R_1,
    R_2, R_3)$ which sum to a rank-1 matrix, and where $A_1, A_2$ and $B_1, B_2$
    are both in general position. We claim that there is only a constant number
    options for $R_3 = A_3 \times B_3$ such that $\rk(R_1 \oplus R_2 \oplus R_3) = 1$.
    Let us first rule out the case where $A_3 \supseteq A_1 \cup A_2$ or $B_3
    \supseteq B_1 \cup B_2$. Let us take representatives $i,j,k$ from $A_1
    \setminus A_2$, $A_1 \cap A_2$, $A_2 \setminus A_1$ respectively, and
    $i',j',k'$ from $B_1 \setminus B_2$, $B_1 \cap B_2$, $B_2 \setminus B_1$
    respectively. The submatrix of $R_1 \oplus R_2$ on rows $\{i,j,k\}$ and
    columns $\{i',j',k'\}$ is
    \[
        P \define (R_1 \oplus R_2)|_{\{i,j,k\} \times \{i',j',k'\}} = J_3 \oplus I_3,
    \]
    where $J_3$ is $3\times 3$ all-$1$ matrix and $I_3$ is the $3\times 3$
    identity matrix. Then if $A_3 \supseteq \{i,j,k\}$ then $P \oplus
    R_3|_{\{i,j,k\} \times \{i',j',k'\}}$ has either two distinct columns from
    $J_3 \oplus I_3$ or two distinct columns from $I_3$, in both cases $\rk(R_1
    \oplus R_2 \oplus R_3) \ge 2$. The same argument shows that $B_3
    \not\supseteq \{i',j',k'\}$. 

    Then there are a row and a column that are nonzero in $R_1 \oplus R_2$ and
    zero in $R_3$. Suppose that their values in $R_1 \oplus R_2$ are $a \in
    \{0,1\}^n$ and $b \in \{0,1\}^n$, respectively. Since $\rk(R_1 \oplus R_2
    \oplus R_3) = 1$, any two nonzero rows of $R_1 \oplus R_2 \oplus R_3$ must be equal,
    and any two nonzero columns must also be equal. Since $a, b$ are nonzero,
    this requires $R_1 \oplus R_2 \oplus R_3 = a \cdot b^T$. For fixed $R_1, R_2$, there are only three
    possible values of $a$ and $b$, so there are at most $9$ options for the
    value of $R_3$. 
    Therefore there are at most $9 \cdot (4^n)^2$ triples $(R_1, R_2, R_3)$ in this case.
\end{proof}

\section{Combinatorial Proof}

We prove a lower bound on the non-deterministic \textsc{Equality} oracle
cost of $\rkone_n$. The non-deterministic \textsc{Equality} oracle cost of any
matrix $M \in \zo^{N \times N}$ is denoted $\ND^\EQ(M)$. It is defined as the
minimum value $t$ such that there are numbers $m, d$ with $m + d = t$, where
there exist matrices $M_1, \dotsc, M_{2^m} \in \zo^{N \times N}$ satisfying:
\begin{enumerate}[noitemsep]
    \item $\DEQ(M_i) \leq d$ for all $i \in [2^m]$;
    \item If $M(x,y) = 1$ then $\exists i \in [2^m]$, $M_i(x,y) = 1$; and
    \item If $M(x,y) = 0$ then $\forall i \in [2^m]$, $M_i(x,y) = 0$.
\end{enumerate}
By definition, $\ND^\EQ(M) \leq \DEQ(M)$. To prove a lower bound on $\ND^\EQ$,
we require the \emph{blocky cover number} defined in \cite{PSS23}.
\begin{definition}[Blocky Cover]
A matrix $B \in \zo^{N \times N}$ is called \emph{blocky} if it is obtained from an identity matrix $I_{m,m}$
with $m \leq N$ by duplicating rows or columns, permuting rows or columns, or adding all-0 rows or columns.
Let $M \in \zo^{N \times N}$ be a boolean matrix. The \emph{blocky cover number} $\bc(M)$ is the
minimum number $r$ such that there exist $r$ blocky matrices $B_1, \dotsc, B_r \in \zo^{N \times N}$
with $M = \bigvee_{i=1}^r B_i$, where $\bigvee$ denotes entrywise OR.
\end{definition}

Blocky covers are related to non-deterministic \textsc{Equality} oracle cost by the following inequality from \cite{PSS23}.  For all $A \in \zo^{N \times N}$,
\begin{equation}
    \ND^\EQ(A) \leq \log \bc(A) \leq O( \ND^\EQ(A) \cdot \log\log N ) .
\end{equation}

In \cref{section:fractional-cover}, we improve the upper bound in this inequality:

\begin{boxlemma}
\label{lemma:bc-ndeq}
For any $A \in \zo^{N \times N}$,
\[
  \ND^\EQ(A) \leq \log \bc(A) \leq \ND^\EQ(A) + O(\log\log N) .
\]
\end{boxlemma}

We may now improve \cref{thm:main} to $\ND^\EQ(\rkone_n) = \Omega(n)$ with the
following bound on blocky cover number.

\begin{boxtheorem}
\label{thm:bc}
     $\bc(\rkone_n) = 2^{\Omega(n)}$. As a consequence, $\ND^\EQ(\rkone_n) = \Omega(n)$.
\end{boxtheorem}
\begin{proof}
In this proof, write $M^{-1}(1)$ for the set of 1-valued entries in a boolean matrix $M$, and write
$|M| = |M^{-1}(1)|$ for the number of 1-valued entries.  We use standard notation $\H(\bm Z)$ for
the binary entropy of a random variable $\bm Z$, so $\H(\bm Z) \define - \sum_{z \in \cZ} \Pr[ \bm
Z = z ] \log \Pr[ \bm Z = z ]$.
   
   Recall from \cref{eq:frobenius} that $|\rkone_n| = 2^{n^2} \cdot 2^{2n}$. A
   rectangle $R = U \times V$ is \emph{1-chromatic} if, for all $(u,v) \in U
   \times V$, $\rkone_n(u,v) = 1$. Let $\bc(\rkone_n) = K$, so $\rkone_n =
   \bigvee_{i \in [K]} B_i$, where the $B_i$ are blocky matrices, which can be
   written as a disjoint union of 1-chromatic rectangles $B_i = \bigvee_{j \in
   [b_j]} R_{i,j}$ where $R_{i,j} = U_{i,j} \times V_{i,j}$ is a 1-chromatic
   rectangle in $B_i$.

\begin{claim}
   \label{cl:large-rectangle}
   There exists a 1-chromatic rectangle $R = U \times V$ with $|U|, |V| \geq 2^{2n}/(2K)$.
\end{claim}
\begin{proof}[Proof of claim]
For the sake of contradiction, assume each 1-chromatic rectangle $R = U \times
V$ has $|U| < 2^{2n}/(2K)$, in which case we call the rectangle \emph{short}, or $|V| <
2^{2n}/(2K)$, in which case we call it \emph{narrow}.
Then by definition of $K$, we obtain a contradiction as follows:
\begin{align*}
    2^{n^2} \cdot 2^{2n}
    &= |\rkone_n|
    \leq \sum_{i=1}^K |B_i| 
    = \sum_{i=1}^K \sum_{j=1}^{b_i} |U_{i,j}| \cdot |V_{i,j}| \\
    &< \sum_{i=1}^K \left( \sum_{j \in [b_i], R_{i,j} \text{ short}} |V_{i,j}| \cdot 2^{2n}/(2K)
                         + \sum_{j \in [b_i], R_{i,j} \text{ narrow}} |U_{i,j}| \cdot 2^{2n}/(2K) \right) \\
    &\leq \tfrac{1}{2} 2^{2n} \max_{i \in [K]} \left( \sum_{j \in [b_i]} |V_{i,j}|
                                    +    \sum_{j \in [b_i]} |U_{i,j}| \right) \\
    &\leq \tfrac{1}{2} 2^{2n} \max_{i \in [K]} \left( 2^{n^2} + 2^{n^2} \right)
    = 2^{2n} \cdot 2^{n^2} . \qedhere
\end{align*}
\end{proof}

\begin{figure}
       \centering

       \definecolor{rectanglecolor}{RGB}{77,98,152}
 \begin{tikzpicture}[scale=0.5]
        \draw[line width=0.7pt] (-0.1,-0.1) rectangle (10.1,10.1);

        \filldraw[fill=rectanglecolor!30, draw=rectanglecolor, line width=0.4pt]
            (0,10) -- (3,10) .. controls (9,5) and (9,6) .. (10,2) --
            (10,2) -- (10,0) -- (8,0) .. controls (4,7) and (2,7) .. (0,9) -- (0,10) -- cycle;
        \node[line width=0.4pt, right] at (3.7,9.5) {$\rkone^{-1}(1)$};

        \fill[rectanglecolor!20] (0.4,10.5) rectangle (7.6,11.5);
        \fill[rectanglecolor!20] (10.5,1.0) rectangle (11.55,9.2);

        \filldraw[pattern={Lines[angle=0,distance=1pt,line width=0.4pt]}, pattern color=rectanglecolor,
                  draw=rectanglecolor, line width=0.4pt]
            (0.5,9.2) rectangle (2.0,9.8);
        \filldraw[pattern={Lines[angle=90,distance=1pt,line width=0.4pt]}, pattern color=rectanglecolor,
                  draw=rectanglecolor, line width=0.4pt]
            (2.4,8.0) rectangle (3.0,9.1);
        \filldraw[pattern={Lines[angle=90,distance=1pt,line width=0.4pt]}, pattern color=rectanglecolor,
                  draw=rectanglecolor, line width=0.4pt]
            (3.825,6.4) rectangle (4.675,7.9);
        \filldraw[pattern={Lines[angle=0,distance=1pt,line width=0.4pt]}, pattern color=rectanglecolor,
                  draw=rectanglecolor, line width=0.4pt]
            (5.2,5.5) rectangle (6.4,6.3);
        \filldraw[pattern={Lines[angle=0,distance=1pt,line width=0.4pt]}, pattern color=rectanglecolor,
                  draw=rectanglecolor, line width=0.4pt]
            (6.5,4.7) rectangle (7.5,5.4);
        \filldraw[pattern={Lines[angle=90,distance=1pt,line width=0.4pt]}, pattern color=rectanglecolor,
                  draw=rectanglecolor, line width=0.4pt]
            (7.8,4.3) rectangle (8.5,1.1);

        \filldraw[pattern={Lines[angle=0,distance=1pt,line width=0.4pt]}, pattern color=rectanglecolor,
                  draw=rectanglecolor, line width=0.4pt]
            (0.5,10.6) rectangle (2.0,11.2);
        \filldraw[pattern={Lines[angle=0,distance=1pt,line width=0.4pt]}, pattern color=rectanglecolor,
                  draw=rectanglecolor, line width=0.4pt]
            (5.2,10.6) rectangle (6.4,11.4);
        \filldraw[pattern={Lines[angle=0,distance=1pt,line width=0.4pt]}, pattern color=rectanglecolor,
                  draw=rectanglecolor, line width=0.4pt]
            (6.5,10.6) rectangle (7.5,11.3);

        \filldraw[pattern={Lines[angle=90,distance=1pt,line width=0.4pt]}, pattern color=rectanglecolor,
                  draw=rectanglecolor, line width=0.4pt]
            (10.6,8.0) rectangle (11.2,9.1);
        \filldraw[pattern={Lines[angle=90,distance=1pt,line width=0.4pt]}, pattern color=rectanglecolor,
                  draw=rectanglecolor, line width=0.4pt]
            (10.6,6.4) rectangle (11.45,7.9);
        \filldraw[pattern={Lines[angle=90,distance=1pt,line width=0.4pt]}, pattern color=rectanglecolor,
                  draw=rectanglecolor, line width=0.4pt]
            (10.6,1.1) rectangle (11.3,4.3);

        \draw[line width=0.4pt, color=rectanglecolor, dotted]
            (0.5,9.8) -- (0.5,10.6) (2.0,9.8) -- (2.0,10.6)
            (5.2,6.3) -- (5.2,10.6) (6.4,6.3) -- (6.4,11.4)
            (6.5,5.4) -- (6.5,10.6) (7.5,5.4) -- (7.5,11.3)
            (3.0,8.0) -- (10.6,8.0) (3.0,9.1) -- (10.6,9.1)
            (4.675,6.4) -- (10.6,6.4) (4.675,7.9) -- (10.6,7.9)
            (8.5,1.1) -- (10.6,1.1) (8.5,4.3) -- (10.6,4.3);

    \end{tikzpicture}
   \caption{Illustration of \cref{cl:large-rectangle}. The rectangles represent the $1$-entries of a blocky matrix $B$ with {$B^{-1}(1) \subseteq \rkone^{-1}(1)$}. Short rectangles are filled with horizontal lines and narrow ones are filled with the vertical lines. We then can conclude that the total area of each type of the rectangles is small. }
       \label{fig:blocky-inside}
   \end{figure}
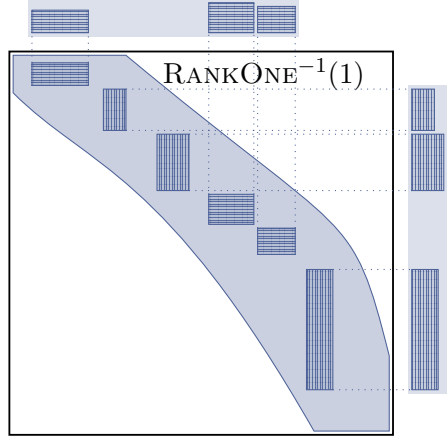

   Suppose that $K < 2^{n/1000}$. Then \cref{cl:large-rectangle} implies that
   there exists a 1-chromatic rectangle $R = U \times V$ with $|U|, |V| \geq
   2^{n(2-1/100)}$. This contradicts the next claim:

   \begin{claim}
       Let $R = U \times V$ be a 1-chromatic rectangle. Then $\min(|U|, |V|)
       < 2^{\alpha n}$ for $\alpha = 2-10^{-2}$.
   \end{claim}
    \begin{proof}[Proof of claim]
    Suppose for the sake of contradiction that $|U|, |V| \geq 2^{\alpha n}$.
    Let $u \in U$ and $v \in V$ be arbitrary, and let  $\bm A \sim U$, $\bm B
    \sim V$ be uniformly random, so that $\bm A, \bm B \in \zo^{n \times n}$.
    Since $R$ is 1-chromatic, $\bm A \oplus v$ and $\bm B \oplus u$ are
    always rank-1 matrices, so we may write $\bm A \oplus v$ as an outer product
    $\bm{\ell}^A \cdot (\bm{r}^A)^T$ where $\bm{\ell}^A, \bm{r}^A \in \zo^n$,
    and similarly $\bm{B} \oplus u \define \bm{\ell}^B (\bm{r}^B)^T$. Then
    $\H(\bm{\ell}^A, \bm{r}^A), \H(\bm{\ell}^B, \bm{r}^B) \ge \alpha n$. By the
    chain rule for entropy, $\sum_{i \text{ odd}} \H(\bm{\ell}^A_i, \bm{r}^A_i,
    \bm{\ell}^A_{i+1}, \bm{r}^A_{i+1}) \ge n(2-10^{-2})$, therefore for at least
    $2/3$-fraction of odd $i \in [n]$, $\H(\bm{\ell}^A_i, \bm{r}^A_i,
    \bm{\ell}^A_{i+1}, \bm{r}^A_{i+1}) \ge 2 - 3\cdot 10^{-2}$. Similarly, for
    at least $2/3$-fraction of odd $i \in [n]$, $\H(\bm{\ell}^B_i, \bm{r}^B_i,
    \bm{\ell}^B_{i+1}, \bm{r}^B_{i+1}) \ge 2 - 3 \cdot 10^{-2}$. Then for
    $1/3$-fraction of odd coordinates $i$, both inequalities are satisfied.
    Suppose without loss of generality that this is the case for $i = 1$.
    
   Then $\H(\bm{\ell}^A_1, \bm{r}^A_1,\bm{\ell}^A_2, \bm{r}^A_2) \geq
   2(2-10^{-2}) > \log 15$, hence $|\supp(\bm{\ell}^A_1, \bm{r}^A_1,
   \bm{\ell}^A_2, \bm{r}^A_2)| = 16$, and similarly $|\supp(\bm{\ell}^B_1,
   \bm{r}^B_1, \bm{\ell}^B_2, \bm{r}^B_2)| = 16$. Thus $(\bm{A} \oplus v)|_{[2]
   \times [2]}$ and $(u \oplus \bm{B})|_{[2] \times [2]}$ can be arbitrary
   rank-$1$ matrices, so in particular there exists $(a,b) \in \supp(\bm{A})
   \times \supp(\bm{B})$ such that $(a\oplus b)|_{[2] \times [2]}$ is the
   identity matrix: take $a$ and $b$ such that $(a\oplus v)|_{[2] \times [2]}$
   and $(u \oplus b)|_{[2] \times [2]}$ form the decomposition of $(v\oplus
   u)|_{[2] \times [2]} \oplus I_2$ as a sum of rank-$1$ matrices. Hence $\rk(a
   \oplus b) \ge 2$, a contradiction.
   \end{proof}
   This concludes the proof of the theorem.
   \end{proof}

\section{Consequences}
\label{section:consequences}

We conclude the paper with some corollaries of our proofs, as mentioned in the introduction.

\subsection{Constant-Cost Communication vs. $k$-Hamming Distance}

We will prove \cref{cor:main}, which shows that constant-cost randomized
communication does not reduce to \textsc{$k$-Hamming Distance}. 

\begin{remark}
\label{rem:query-size}
We must clarify a subtlety
in the definition of $\D^\cQ(P)$, the cost of a deterministic communication protocol with access
to an oracle for problem $\cQ$.
For the purpose of studying the class $\BPP_0$, the natural definition of
$\D^\cQ$ is as follows (see \eg \cite{HWZ22,FHHH24,FGHH25}). $\cQ$ is a family
of boolean matrices, \ie a communication problem, containing matrices $Q \in
\zo^{N \times N}$ for a variety of input sizes $N$. $\D^\cQ(P)$ is the minimum
cost of a deterministic protocol computing $P \in \zo^{N \times N}$, where on
input $x,y$, the two parties Alice and Bob in each round may construct inputs
$a(x)$ and $b(y)$ of \emph{arbitrary size}, and query the oracle on $Q(a(x),
b(y))$, where $Q \in \cQ$ is any instance of problem $\cQ$. The difference
between this definition and the standard definition of oracle protocols is that
the query inputs $a(x), b(y)$ do not have bounded size. This is natural because
for $\cQ \in \BPP_0$, the randomized communication cost is independent of input
size.
\end{remark}

Recent work \cite{FHHH24} showed that there is no complete problem for $\BPP_0$:
there is no problem $\cQ \in \BPP_0$ such that $\D^\cQ(P) = O(1)$ for all
problems $P \in \BPP_0$. The reason is that, for every $\cQ \in \BPP_0$, there
exists a constant $k$ such that $\D^\cQ(\HD{k}{n}) = \omega(1)$. This raised the
question of whether the infinite hierarchy of \textsc{$k$-Hamming Distance}
problems $\HD{1}{}, \HD{2}{}, \dotsm$ is complete, meaning that for every
problem $\cP \in \BPP_0$ there exists a constant $k$ such that
$\D^{\HD{k}{}}(P_N) = O(1)$. The question was answered in \cite{FGHH25}, whose
main result was:

\begin{theorem}[\cite{FGHH25}]
There exists a communication problem $F_n \colon \zo^n \times \zo^n \to \zo$ such that
$\R(F_n) = O(1)$, yet for every constant $k$, $\D^{\HD{k}{}}(F_n) = \omega(1)$.
\end{theorem}

The $\omega(1)$ function is not given explicitly and is the result of a lengthy
Ramsey-theoretic proof. We give a simpler proof (using a different function $F$)
with a quantitative bound.

\begin{proof}[Proof of \cref{cor:main}]
We prove that, for every constant $k$, $\D^{\HD{k}{}}(\rkone_n) = \Omega(n / \log n)$.
Given a deterministic protocol computing $\rkone_n$ using $\HD{k}{}$ queries, we may replace each
query $Q$ with a protocol computing $Q$ using \textsc{Equality} queries. Then
\[
    \DEQ(\rkone_n) \leq \D^{\HD{k}{}}(\rkone_n) \cdot \max_Q \DEQ(Q) ,
\]
where the maximum is over all $2^{n^2} \times 2^{n^2}$ matrices $Q$ which
represent oracle queries to $\HD{k}{}$; formally, these matrices $Q$ are the
ones which are obtained by taking any submatrix of $\HD{k}{n'}$ for arbitrarily
large $n'$ (due to \cref{rem:query-size}), and then duplicating rows or columns
to ensure the matrix is the same size as $\rkone_n$. We now require a bound on
$\DEQ(Q)$ for arbitrary $2^{n^2} \times 2^{n^2}$ submatrices of $\HD{k}{n'}$
where $n'$ can be arbitrarily large. This was proved in \cite{FHHH24}:

\begin{theorem}[{\cite[Proposition 4.1]{FHHH24}}]
\label{thm:fhhh-khd-deq}
For any $k$ and any $n, N \in \bN$ satisfying $N \leq 2^n$, let $M$ be an $N \times N$ submatrix
of $\HD{k}{n} \in \zo^{2^n \times 2^n}$. Then $\DEQ(M) \leq O(k \log\log N)$.
\end{theorem}

So, by \cref{thm:fhhh-khd-deq},
\[
    \DEQ(\rkone_n) \leq O\left( \D^{\HD{k}{}}(\rkone_n) \cdot k \log\log(2^{n^2}) \right) .
\]
By \cref{thm:main}, we conclude that $\D^{\HD{k}{}}(\rkone_n) = \Omega\left(\frac{n}{k\log n}\right)$.
\end{proof}

\subsection{Parity Decision Trees and Exact vs.~Approximate Spectral Norm}
\label{section:pdt}

We state some consequences for parity decision trees and exact vs.~approximate spectral norms.
The Fourier coefficients of a function
$f\colon \{0,1\}^n \to \mathbb{R}$ are defined on subsets $S \subseteq [n]$ by
\[
\hat{f}(S) \define \mathbb{E}_{\bm{x} \sim \{0,1\}^n}[f(\bm{x}) (-1)^{\sum_{i \in S} \bm{x}_i}].
\]
The Fourier norm is $\|\hat{f}\|_1 \define \sum_{S \subseteq [n]}
|\hat{f}(S)|$. For an XOR problem $F(x,y) = f(x \oplus y)$, it is equivalent to the $\gamma_2$-norm:
\begin{equation}
    \gamma_2(F) = \|\hat f\|_1 
    \tag{\cite[Corollary~3.9]{HHH23}}
\end{equation}
For $\epsilon \in (0,1)$, the approximate spectral norm is defined as
\[
    \| \widehat f \|_{1,\epsilon} \define \inf_g \|\widehat g\|_1
\]
where the infimum is over functions $g \colon \zo^n \to \mathbb{R}$
which satisfy $\|f-g\|_\infty < \epsilon$.

We can now state the corollaries. The first is an exponential improvement over the separation given by \cite[Lemma~3]{CHZZ24}.
\begin{boxcorollary}
\label{cor:approx-spectral-norm}
There exists a function $f \colon \zo^{n \times n} \to \zo$ with approximate spectral norm
$\|\widehat f\|_{1, 1/3} = O(1)$ but exact spectral norm $\|\widehat f\|_1 = 2^{\Omega(n)}$.
\end{boxcorollary}
We remark that for some constants $\epsilon$ and $\ell$ the bound $\|\widehat{f}\|_{1,\epsilon} \le \ell$ \cite[Theorem~2.5]{Sanders2019} (known as quantitative Cohen idempotence theorem) implies that $\|\widehat{f}\|_1 = O(1)$, so this corollary (as well as \cite[Lemma~3]{CHZZ24}) gives a barrier for improving the dependency between $\ell$ and $\epsilon$ in this result.

The second corollary answers \cite[Question 8]{CHHNPS25}.
\begin{boxcorollary}
\label{cor:rpdt}
There exists a function $f \colon \zo^{n\times n} \to \zo$ with randomized parity decision tree
size $O(1)$ and deterministic parity decision tree size $2^{\Omega(n)}$.
\end{boxcorollary}

The separating function $f \colon \zo^{n\times n} \to \zo$ in both cases is defined by $f(A) = 1$ iff $A \in \zo^{n \times n}$ has $\rk(A)\leq 1$, so $\rkone_n(x,y) = f(x \oplus y)$.

\begin{proof}[Proof of \cref{cor:approx-spectral-norm}]
    By \cite[Proposition 4.3]{HHH23} we get that $\|\widehat{f}\|_{1,1/3} = O(1)$ since $1/3$-error randomized communication cost of $\rkone$ is $O(1)$ by \cref{thm:rpdt-upper-bound}. On the other hand, by \cite[Corollary~3.9]{HHH23} combined with \cref{thm:gamma2} we get $\|\widehat{f}\|_1 = 2^{\Omega(n)}$ as required.
\end{proof}

\begin{proof}[Proof of \cref{cor:rpdt}]
The upper bound on randomized parity decision tree size of $f$
was proved in  \cref{thm:rpdt-upper-bound}.
If $f$ has a parity decision tree (PDT) of size $s$, then $f(x) = \sum_{i \in [s]} {A_i}(x)$ where
$A_i$ are characteristic functions of the affine spaces corresponding to
$1$-labeled leaves of the PDT. It is well known that $\|\widehat{A_i}\|_1\le
1$. Then $\|\hat{f}\|_1 \le \sum_{i \in [s]} \|\widehat{{A_i}}\|_1 \le s$. By
\cref{thm:gamma2},
\[
    2^{\Omega(n)} \leq \gamma_2(\rkone_n) = \|\hat f\|_1 \leq s . \qedhere
\]
\end{proof}

\subsection{Non-Deterministic Lower Bounds}

The paper \cite{CLV19} showed that $\BPP \not\subseteq \P^\EQ$ using the
\textsc{Integer Inner Product} function, defined as $\IIP_3^n\colon [-2^n,2^n]^3
\times [-2^n,2^n]^3 \to \{0,1\}$, where $\IIP_3^n(x,y) = 1$ iff $\sum_{i \in
[3]} x_i y_i = 0$. Later, \cite{PSS23,CHHNPS25} improved this separation to
$\BPP \not\subseteq \NP^\EQ$ using $\IIP_3$ in the latter work (and the
higher-dimensional $\IIP_6$ in the prior work). \cite{CHHNPS25} prove a lower
bound of $\ND^\EQ(\IIP_3^n) = \Omega(n / \log n)$ using \cref{eq:fbc-product}
and asking whether the $\log n$ factor could be removed. With our improved
\cref{lemma:bc-ndeq} we can remove this log factor.

\newcommand{\maxrect}{\mathrm{maxrect}}

We need the notion of $\maxrect(\cdot)$ from \cite{PSS23}. For a matrix $A \in
\{0,1\}^{N \times N}$ let $\alpha(A)$ be the number of $1$-entries in $A$ and
let $\beta(A)$ be the area of the largest $1$-chromatic rectangle in $A$.
Then
\[
\maxrect(A) \define \frac{\alpha(A)}{N \sqrt{\beta(A)}}.
\]
The following two theorems, together with the fact that $\bc(\cdot)$ does not increase when taking submatrices,
imply $\bc(\IIP_3^n) = 2^{\Omega(n)}$:
\begin{theorem}[Theorem~5 in \cite{CHHNPS25}]
    $\IIP_3^n$ has a submatrix $A$ such that $\maxrect(A) = 2^{\Omega(n)}$.
\end{theorem}
\begin{theorem}[Theorem~29 in \cite{PSS23}]
    For every $A \in \{0,1\}^{N \times N}$ we have $\bc(A) \ge \Omega(\maxrect(A))$.
\end{theorem}
Then, by \cref{lemma:bc-ndeq} (and the trivial bound $\DEQ(\IIP_3^n) = O(n)$):
\begin{boxcorollary}
    $\ND^{\EQ}(\IIP_3^n) = \Theta(n)$. 
\end{boxcorollary}

\appendix

\section{Appendix: Proof of \texorpdfstring{\cref{lemma:bc-ndeq}}{\ref{lemma:bc-ndeq}}}
\label{section:fractional-cover}

Let us now prove \cref{lemma:bc-ndeq}.  Let $A \in \zo^{N \times N}$. Our goal is to show that
\[
  \log \bc(A) \leq  \ND^\EQ(A) + O( \log\log N ).
\]
To prove this, we will use the fractional blocky cover instead of the blocky cover:

\newcommand{\fbc}{\mathrm{fbc}}
\begin{definition}[Fractional Blocky Cover]
For any $A \in \zo^{N \times N}$, a \emph{fractional blocky cover} of $A$ is a choice of values $m
\in \bN$, $\lambda_1, \dotsc, \lambda_m \geq 0$, and blocky matrices $B_1, \dotsc, B_m$ such that
\begin{align*}
  A(x,y) = 0 &\implies \sum_{i=1}^m \lambda_i B_i(x,y) = 0 
  &A(x,y) = 1 &\implies \sum_{i=1}^m \lambda_i B_i(x,y) \geq 1 .
\end{align*}
We write $\fbc(A)$ for the minimum value of $\sum_i \lambda_i$ over all fractional blocky covers of
$A$.
\end{definition}

We require the following properties of the fractional blocky cover:
\begin{proposition}[Properties of FBC]
\label{prop:properties-fbc}
Let $P, Q \in \zo^{N \times N}$ be any boolean matrices and let $J$ be the all-1s matrix. Then
\begin{enumerate}[noitemsep]
\item For any blocky matrix $B$, $\fbc(J-B) \leq 4$.
\item $\fbc(P \wedge Q) \leq \fbc(P) \cdot \fbc(Q)$.
\item $\fbc(P \vee Q) \leq \fbc(P) + \fbc(Q)$.
\item $\bc(P) \leq O( \fbc(P) \cdot \log N)$. 
\end{enumerate}
\end{proposition}
We prove these properties below. First let us complete the proof of \cref{lemma:bc-ndeq}.

\begin{proof}[Proof of \cref{lemma:bc-ndeq}]
Let $A \in \zo^{N \times N}$.  Due to property (4) of \cref{prop:properties-fbc}, it suffices to
show $\log\fbc(A) = O(\ND^\EQ(A))$. Let $d, m \in \bN$ be such that $d+m = \ND^\EQ(A)$ and there exist
$2^m$ deterministic \textsc{Equality}-oracle protocols $T_i$ of depth at most $d$ such that
\[
  \forall x,y \in [N] \;\colon\qquad A(x,y) = \bigvee_{i=1}^{2^m} T_i(x,y) .
\]
By property (3) of \cref{prop:properties-fbc}, $\fbc(A) \leq \sum_{i=1}^{2^m} \fbc(M_i)$ where $M_i
\in \zo^{N \times N}$ is the matrix computed by protocol $T_i$. It now suffices to show that
$\fbc(M_i) \leq 5^d$ for every $i$, so that $\log\fbc(A) \leq \log(2^m 5^d) < 3 \cdot \ND^\EQ(A)$.

Fix any $T = T_i$, which is a decision tree with each inner node $v$ associated with a blocky matrix
$B_v$. Without loss of generality we may assume the depth is exactly $d$ and that the tree is full.
For each node $v$, let $M_v \in \zo^{N \times N}$ be the matrix computed by the subtree rooted at
$v$, and write $\mathrm{depth}(v)$ for the depth of this subtree. We prove by induction that
$\fbc(M_v) \leq 5^{\mathrm{depth}(v)}$. If $\mathrm{depth}(v) = 1$, then $M_v$ is either a blocky
matrix $B$ or its complement $J-B$, so by property (1) of \cref{prop:properties-fbc}, $\fbc(M_v)
\leq 4$. For depth $d' > 1$, let $\ell, r$ be the left and right child of $v$. Then
\[
  M_v = (B_v \wedge M_\ell) \vee ((J-B_v) \wedge M_r ) ,
\]
so by properties (1), (2), and (3) of \cref{prop:properties-fbc},
\begin{align*}
  \fbc(M_v) &\leq \fbc(B_v) \fbc(M_\ell) + \fbc(J-B_v) \fbc(M_r) \leq \fbc(M_\ell) + 4\fbc(M_r) \\
    &\leq 5^{d'-1} + 4 \cdot 5^{d'-1} = 5^{d'} ,
\end{align*}
with the last inequality by induction. This completes the proof of the lemma.
\end{proof}

Finally, we establish the properties of fractional blocky covers that we have used above.

\begin{proof}[Proof of \cref{prop:properties-fbc}]
\begin{claim}
Let $J$ be the all-1s matrix and let $B$ be any blocky matrix. Then $\fbc(J-B) \leq 4$.
\end{claim}
\begin{proof}
It suffices to prove the claim for the identity matrix $B = I_{N \times N}$, since duplicating rows
and columns does not change the value of $\fbc$.  The claim follows from the fact that the
\textsc{Equality} communication problem has a constant-cost one-sided error randomized protocol.
That is, define the following probability distribution over rectangles $R \subseteq [N] \times
[N]$. We choose $\bm R$ by including each $x \in [N]$ in a set $\bm X \subseteq [N]$ independently
with probability $1/2$, and then take $\bm R = \bm X \times ([N] \setminus \bm X)$. If
$x = y$, then $\Pr[ (x,y) \in \bm R ] = 0$. If $x \neq y$ then $\Pr[ (x,y) \in \bm R ] = 1/4$.
For each rectangle $R$, set $\lambda_R \define 4 \cdot \Pr[ \bm R = R ]$. It follows that
\[
  \fbc(J-I_{N \times N}) \leq \sum_R \lambda_R = 4 . \qedhere
\]
\end{proof}

\begin{claim}
Let $P, Q \in \zo^{N \times N}$. Then $\fbc(P \wedge  Q) \leq \fbc(P) \cdot \fbc(Q)$.
\end{claim}
\begin{proof}[Proof of claim]
Let $\alpha_1, \dotsc, \alpha_p$ and $B_1, \dotsc, B_p$ be a fractional blocky cover of $P$, and let
$\beta_1, \dotsc, \beta_q$ and $B'_1, \dotsc, B'_q$ be a fractional blocky cover of $Q$. For any
$x,y \in [N]$, consider
\begin{equation}
\label{eq:fbc-product}
  \sum_{i,j} \alpha_i \beta_j (B_i \wedge B'_j)(x,y)
  = \left( \sum_{i} \alpha_i B_i(x,y) \right) \cdot \left( \sum_{j} \beta_j B'_j(x,y) \right)
\end{equation}
Note that $B_i \wedge B'_j$ is itself a blocky matrix.
If $P \wedge Q(x,y) = 0$ then $P(x,y) = 0$ or $Q(x,y) = 0$ so \cref{eq:fbc-product} is 0.  If $P
\wedge Q(x,y) = 1$ then $P(x,y) = Q(x,y) = 1$ so \cref{eq:fbc-product} is at least 1. So
\[
  \fbc(P \wedge Q) \leq \sum_{i,j} \alpha_i \beta_j = \fbc(P) \cdot \fbc(Q) . \qedhere
\]
\end{proof}

\begin{claim}
Let $P, Q \in \zo^{N \times N}$. Then $\fbc(P \vee  Q) \leq \fbc(P) + \fbc(Q)$.
\end{claim}
\begin{proof}[Proof of claim]
Let $\alpha_1, \dotsc, \alpha_p$ and $B_1, \dotsc, B_p$ be a fractional blocky cover of $P$, and let
$\beta_1, \dotsc, \beta_q$ and $B'_1, \dotsc, B'_q$ be a fractional blocky cover of $Q$. It is
straightforward to check that the union of these fractional blocky covers is itself a fractional
blocky cover for $P \vee Q$, so that $\fbc(P \vee Q) \leq \sum_i \alpha_i + \sum_j \beta_j$.
\end{proof}

\begin{claim}
Let $P \in \zo^{N \times N}$. Then $\bc(P) \leq O(\fbc(P) \cdot \log N)$.
\end{claim}
\begin{proof}[Proof of claim]
This is by standard randomized rounding. Let $\lambda_1, \dotsc, \lambda_m \geq 0$ and $B_1, \dotsc,
B_m$ be a fractional blocky cover of $P$ with $\sum_\ell \lambda_\ell = \fbc(P)$. Let $\bm M_1,
\dotsc, \bm M_t$ be random blocky matrices chosen independently from the distribution $\Pr[ \bm M_i
= B_j ] = \frac{\lambda_j}{\fbc(P)}$. Fix any $x,y \in [N]$ and note that
\[
  P(x,y) = 0 \implies \Pr[ \exists i \;\colon\; \bm M_i(x,y) = 1 ] = 0
\]
since $B_\ell(x,y) = 0$ for all $\ell \in [m]$. Now suppose $P(x,y) = 1$. Then
$\sum_\ell \lambda_\ell B_\ell(x,y) \geq 1$, so, by independence of each $\bm M_i$,
\[
  \Pr[ \forall i \;\colon\; \bm M_i(x,y) = 0 ] = \Pr[\bm M_1(x,y) = 0]^t = \left(1 -
\frac{1}{\fbc(P)}\right)^t \leq e^{-\frac{t}{\fbc(P)}} .
\]
Setting $t = O(\fbc(P) \log N)$ and using the union bound over at most $N^2$ pairs $x,y \in [N]$
with $P(x,y) = 1$, we conclude that there exists a choice of $t = O(\fbc(P) \log N)$ blocky matrices
which cover the 1-valued entries of $P$.
\end{proof}

This completes the proof of all of the properties from \cref{prop:properties-fbc}.
\end{proof}

\medskip

\section*{Acknowledgments}

Thanks to Kaave Hosseini for suggesting \cref{cor:approx-spectral-norm} and to Tsun-Ming Cheung for
comments on the presentation of this article. Thanks to an anonymous reviewer for pointing out a
simpler proof of \cref{cl:large-rectangle}.

The project is supported by Swiss
State Secretariat for Education, Research, and Innovation (SERI) under contract number MB22.00026.

\DeclareUrlCommand{\Doi}{\urlstyle{sf}}
\renewcommand{\path}[1]{\small\Doi{#1}}
\renewcommand{\url}[1]{\href{#1}{\small\Doi{#1}}}
\bibliographystyle{alphaurl}
\bibliography{bib.bib}

\end{document}